\documentclass[preprint,showpacs,aps]{revtex4}

\usepackage{amsmath}

\begin{document}

\title{Dust acoustic wave in a strongly magnetized pair-dust plasma}
\author{P. K. Shukla and M. Marklund}
\affiliation{Institut f\" ur Theoretische Physik IV, Fakult\"at f\"ur
  Physik und Astronomie, Ruhr--Universit\"at Bochum, D--44780 Bochum,
  Germany} 
\date{Received 14 April 2004}

\begin{abstract}
The existence of the dust acoustic wave (DAW) in a strongly magnetized electron-positron
(pair)-dust plasma is demonstrated. In the DAW, the restoring force comes from
the pressure of  inertialess electrons and positrons, and the dust mass provides the 
inertia. The waves could be of interest in astrophysical settings such as the 
supernovae and pulsars, as well as in cluster explosions by intense laser beams in 
laboratory plasmas. 
\end{abstract}

\maketitle

\section{Introduction}

The dust acoustic wave (DAW) in an unmagnetized dusty plasma was
theoretically predicted by  Rao, Shukla and Yu \cite{a1}, which
revolutionized the field of dusty plasma physics. The phase speed of
the DAW is much smaller (larger) than the electron and ion (dust) thermal
speed. The DAW is observed in several laboratory experiments
\cite{a2,a3,a4,a5}. The properties of the DAW in a magnetized
electron-ion-dust plasma has been discussed in a textbook \cite{a6}.    

In this paper, we consider the dispersive property of the DAW in a 
strongly magnetized pair-dust plasma in which electrons and positrons 
have equal masses. Such a pair-dust plasma can occur in supernovae and 
pulsar environments, as well as in cluster explosions by intense laser  
beams in laboratory experiments. In supernovae and around pulsars, the
production of electron--positron pairs is considerable, and the
magnetic field strengths can also become appreciable, especially close
to the surface of neutron stars ($B_{\text{surf}} \sim
10^{10}-10^{13}$ G). These extreme environments are well-known to host
large quantities of electron--positron plasmas, as well as dust
particles \cite{Tajima-Shibata}. 
Clusters, on the other hand, are laboratory produced bonded atomic
structures, which when irradiated by intense laser 
fields form plasmas on a nano-scale. These plasmas have a life span of 
the order of femto seconds, after which they erupt in a release of highly
energetic electrons and dust particles \cite{clusters,liu}. It is expected
that in the next generation of lasers \cite{mou98} and the free
electron lasers which are under development (see e.g.\ \cite{xfel})
will make it possible to produce electron, positrons and dust
particles due to cluster explosions. Moreover, as the intense laser
pulses interact with the plasma, strong magnetic fields will be
generated, thus creating an environment similar to the one
investigated in this paper. 

The frequency (phase speed) of the DAW in our pair-dust plasma is much 
smaller than the electron/positron gyrofrequency (electron and positron 
thermal speeds). In the dust acoustic wave potential, both electrons and
positrons rapidly thermalize along the magnetic field direction and
establish Boltzmann distributions. Thus, pressures of inertialess
electrons and positrons provide the restoring force, whereas the dust
mass provides the inertia in order for the DAW to exist in a
magnetized pair-dust plasma. The frequency of the DAW in a pair-dust
magnetoplasma is found to be larger than that in an unmagnetized
electron-ion-dust plasma. 

\section{Formulation}

Let consider an electron-positron-dust plasma in an external magnetic field 
$B_0 \hat {\bf z}$, where $B_0$ is the magnitude of the external magnetic 
field and $\hat {\bf z}$ is the unit vector along the $z$ axis. At
equilibrium, we have $e n_{e0}- q_d n_{d0} = e n_{p0}$, where $e$ is magnitude of the 
electron charge, $n_{j0}$ is the unperturbed particle number density of 
the particle species $j$ ($j$ equals $e$ for electrons, $p$ for positrons, 
and $d$ for dust grains which have the charge $q_d$; where $q_d
=-eZ_d$ for negatively charged dust grains, $q_d =eZ_d$ for positively charged 
dust grains, and $Z_d$ is the number of charges on the dust grain surface).
The perpendicular (to $\hat {\bf z}$) component of the electron and
positron fluid velocities in the presence of low-frequency 
(in comparison with $\omega_{c} =eB_0/m c$, where $m$ is the
electron/positron mass and $c$ is the speed of light in vacuum)
electrostatic field ${\bf E} = -\nabla \phi$, where $\phi$ is the
scalar potential, are
%1
\begin{subequations}
\begin{equation}
{\bf v}_{\perp e} \simeq \frac{c}{B_0}\hat {\bf z} \times \nabla \phi
- \frac{c T_e}{eB_0 n_{e0}} \hat {\bf z} \times \nabla n_{e1} 
+\frac{c}{B_0 \omega_{c}}\partial_t \nabla_\perp \phi,
\end{equation}
and
%2
\begin{equation}
{\bf v}_{\perp p} \simeq\frac{c}{B_0}\hat {\bf z} \times \nabla \phi 
+ \frac{c T_p }{eB_0 n_{p0}}\hat {\bf z}\times \nabla n_{p1} -
\frac{c}{B_0 \omega_{c}}\partial_t \nabla_\perp \phi,
\end{equation}
\label{eq:velocity-perp}
\end{subequations}
respectively, 
where $T_e (T_p)$ is the electron (positron) temperature and $n_{j1}
(\ll n_{j0}$ is a small density perturbation in the equilibrium value $n_{j0}$.
We have denoted $\partial_t =\partial /\partial t$.  The parallel component of 
the electron and positron  fluid velocities are obtained from
%3
\begin{subequations}
\begin{equation}
\partial_t v_{ez} =\frac{e}{m}\partial_z \left(\phi- \frac{T_e n_{e1}}{en_{e0}}\right),
\end{equation}
and
%4
\begin{equation}
\partial_t v_{pz} =-\frac{ e}{m}\partial_z \left(\phi + \frac{T_p n_{p1}}{en_{p0}}\right).
\end{equation}
\label{eq:velocity-par}
\end{subequations}

Inserting Eqs.\ (\ref{eq:velocity-perp}) and (\ref{eq:velocity-par})
into the electron and positron continuity 
equations and assuming that $|\partial_t| \ll V_{Te, Tp} |\partial_z|,
\, \omega_c |\partial_z|$, we obtain 
%5
\begin{equation}
n_{e1} \simeq \frac{n_{e0} e \phi}{T_e} \quad \text{and} \quad 
n_{p1} \simeq -\frac{n_{p0} e \phi}{T_p} ,
\label{eq:boltzmann}
\end{equation}
which indicate that both electrons and positrons will thermalize
rapidly along the magnetic field direction $\hat{\bf z}$, and
establish Boltzmann distributions. Here $V_{Te}
=(T_e/m)^{1/2}$ and $V_{Tp} =(T_p/m)^{1/2}$ are the electron and
positron thermal speeds, respectively. 

We are interested in the DAW whose frequency is much larger than the
dust gyrofrequency. Thus, charged dust grains are unmagnetized and their
dynamics is governed by the dust continuity and momentum equations
 
%7
\begin{equation}
\partial_t n_{d1} + n_{d0}\nabla \cdot {\bf v}_d =0.
\label{eq:continuity}
\end{equation}
and 
%8
\begin{equation}
\partial_t {\bf v}_d = - \frac{q_d}{m_d}\nabla \phi,
\label{eq:momentum}
\end{equation}
where $m_d$ is the dust mass. Equations
(\ref{eq:boltzmann})--(\ref{eq:momentum}) are closed by means of 
Poisson's equation
%9
\begin{equation}
\nabla^2 \phi =4\pi (e n_{e1} - q_d n_{d1}- en_{p1}).
\label{eq:poisson}
\end{equation}
which takes the form
\begin{equation}
  (\nabla^2 - k_D^2)\phi = -4\pi q_d n_{d1} , 
\label{eq:poisson2}
\end{equation}
using Eq.\ (\ref{eq:boltzmann}). Here $k_D^2
=(1/\lambda_{De}^2)+(1/\lambda_{Dp}^2)$, with $\lambda_{De}=(T_e/4\pi
n_{e0}e^2)^{1/2}$ and $\lambda_{Dp} = (T_p/4\pi n_{p0}e^2)^{1/2}$
being the electron and positron Debye radius, respectively. 

Combining (\ref{eq:continuity}) and (\ref{eq:momentum}) we have
%10
\begin{equation}
\partial_t^2 n_{d1} =\frac{n_{d0}q_d}{m_d}\nabla^2 \phi
\label{eq:wave}
\end{equation}
Substituting for the Laplacian of the potential using Eq.\
(\ref{eq:poisson}), and
eliminating $n_{e,p1}$ by means of (\ref{eq:boltzmann}) we have  
%11
\begin{equation}
(\partial_t^2+ \omega_{pd}^2)n_{d1}  = \frac{\omega_{pd}^2k_D^2}{4\pi
  q_d} \phi ,
\label{eq:wave2}
\end{equation}
where $\omega_{pd}=(4\pi n_{d0} q_d^2/m_d)^{1/2}$ is the dust 
plasma frequency.

Supposing that $\phi$ and $n_{d1}$ is proportional to $\exp(i{\bf
k}\cdot {\bf r} - i  \omega t)$, where ${\bf k}$ and $\omega$ are
the wavevector and frequency, Eqs.\ (\ref{eq:poisson2}) and
(\ref{eq:wave2}) yield
%12
\begin{equation}
\omega = \frac{k C_D} {(1+ k^2 \lambda_D^2)^{1/2}},
\end{equation}
which is the dust acoustic wave frequency in a strongly magnetized
pair plasma.  Here $C_D =\omega_{pd}\lambda_D$ is the dust acoustic speed and
$\lambda_D= \left[ \lambda_{De}\lambda_{Dp}/(\lambda_{De}^2 +
\lambda_{Dp}^2)\right]^{1/2}$ is the  
effective pair plasma Debye radius. We note the resemblance between
the frequency spectrum (12)  
of the DAW in an electron-positron-dust with that DAW in an
unmagnetized electron-ion-dust  
plasma. In the latter with $T_e \gg T_i$ and $n_{i0} > n_{e0}$, we
have $\lambda_D = 
\lambda_{Di} \ll \lambda_{De}$, where $\lambda_{Di}=(T_i/4\pi
n_{i0}e^2)^{1/2}$ is the ion 
Debye radius. On the other hand, in an electron-positron-ion plasma,
we typically have 
$\lambda_D =\lambda_{De}/(1+\sigma)^{1/2}$, where $\sigma=T_e
n_{p0}/T_p n_{e0}$. 
Thus, the dispersion properties of the DAW in a magnetized pair-dust
plasma are  
significantly different from those in an unmagnetized
electron-ion-dust plasma.

\section{Summary}

In summary, we have examined the linear property of intermediate
frequency, viz.  $\omega_{cd} =|q_d| B_0/m_dc \ll \omega \ll \omega_c$, DAW 
in a strongly magnetized pair-dust plasma. The DAWs are supported 
by the pressure of the inertialess electrons and positrons, whereas the
mass of the dust grains provide the inertia. 

The results presented in this paper should be of importance in the
application of the next generation lasers to cluster irradiation, and
the ensuing cluster explosion. These events are likely to produce an
environment in which pair plasmas, strong magnetic fields, and dust
particles coexist. Similarly, in supernovae, the pair plasma will
inevitably contain dust particles, and will be interpenetrated by
magnetic fields, thus creating physical conditions at which the DAW
derived in this paper will occur. 

\acknowledgments
The authors dedicate this paper to Lennart Stenflo on the occasion of
his 65th birthday.

This work was partially supported by the European Commission
(Brussels, Belgium)  
through contract No. HPRN-CT-2001-00e14 for carrying out the task of
the Human Potential Research Training Networks ``Turbulent Boundary
Layer'', as well as by the Deutsche Forschungsgemeinschaft (Bonn,
Germany) through the Sonderforschungsbereich 591 and by DOE Grant No
DE-FG03-97ER54444.

%%%%%% BIBLIOGRAPHY %%%%%%

%%%%%% END OF BIBLIOGRAPHY %%%%%%

\end{document}